**Title: Key principles for workforce upskilling via online learning: a learning analytics study of a professional course in additive manufacturing**

Kylie Peppler*[1], Joey Huang[1], Michael C. Richey[2], Michael Ginda[3], Katy Börner[3], Haden Quinlan[4], A. John Hart[4]

[1] Department of Informatics, University of California, Irvine, California, United States of America

[2] Associate Technical Fellow, The Boeing Company, United States of America

[3] Department of Intelligent Systems Engineering, Luddy School of Informatics, Computing, and Engineering, Indiana University, Bloomington, United States of America

[4] Department of Mechanical Engineering and Center for Additive and Digital Advanced Production Technologies, Massachusetts Institute of Technology, Cambridge, Massachusetts, United States of America

*Corresponding author

E-mail: kpeppler@uci.edu

# Abstract

Effective adoption of online platforms for teaching, learning, and skill development is essential to both academic institutions and workplaces. Adoption of online learning has been abruptly accelerated by COVID-19 pandemic, drawing attention to research on pedagogy and practice for effective online instruction. Moreover, online learning requires a multitude of skills and resources spanning from learning management platforms to interactive assessment tools, combined with multimedia content, presenting challenges to many instructors and organizations. This study focuses on ways that learning sciences and visual learning analytics can be used to design, and to improve, online workforce training in advanced manufacturing. University faculty and industry experts, educational researchers, and specialists in data analysis and visualization collaborated to study the performance of a cohort of 900 professionals enrolled in an online training course focused on additive manufacturing The course was offered through MIT xPRO, MIT Open Learning's professional learning organization, and hosted in a dedicated instance of the massive open online course (MOOC) edX platform. This study combines learning objective analysis and visual learning analytics to examine the relationships among learning trajectories, engagement, and performance. The results demonstrate how visual learning analytics can be used for targeted course modification, and interpretation of learner engagement and performance, such as by more direct mapping of assessments to learning objectives, and to expected and actual time needed to complete each segment of the course. The study also emphasizes broader strategies for course designers and instructors to align course assignments, learning objectives, and assessment measures with learner needs and interests, and argues for a synchronized data infrastructure to facilitate effective just-in-time learning and continuous improvement of online courses.

# Introduction

Effective adoption of online platforms for teaching, learning, and skill development is essential to both academic institutions and workplaces. Adoption of online learning has been abruptly accelerated by COVID-19 pandemic, drawing attention to research on pedagogy and practice for effective online instruction. Meanwhile, educators and trainers are woefully underprepared for the needs of designing for this space, especially for challenging content areas, such as those with specific physical environment needs. What is needed are more systemic approaches to scaffold high-quality online learning outcomes, grounded in the science of how people learn [1] and how to design for online skill development.

Some of the most challenging content areas to adapt concern those requiring hands-on elements and/or three-dimensional visualizations, including Computer-Aided Design (CAD), robotics, and additive manufacturing. These topics also encompass critical future skills for manufacturing and related industries [2], resulting in federal and local government initiatives that coordinate long-term solutions for higher education institutions and companies to face this upskilling challenge [3, 4]. The collective growth of automation and data-driven manufacturing infrastructure foreshadows a disruptive challenge to production processes and human machine interaction and indicates a transformation concerning the future of work. This sociotechnical-networked production system will demand emergent skills for production engineers and more flexibility in the production workforce. As production systems shift to intelligent cognitive agents embedded in the flow of work, all aspects of work will be affected; culture, work methods, organization and spans of control, data systems, division of labor, management practices, and communities must adapt. This disruption impacts human labor, employment and professional competencies (e.g., artificial intelligence and machine learning for robotics, intelligent cognitive assistants) will require a new model for collaboration and human-machine cooperation [5].

The emergence of these new technologies, such as additive manufacturing, and their convergence with other elements within complex manufacturing systems, results in an information gap between employers, employees, and educational institutions. As these technologies mature, the knowledge gap between emergent digital manufacturing and data analytic skills needed by industry grow further apart. Even though employers presumably know which skills they value in an employee, it is challenging for workers themselves and educational institutions to match the pace of change required to meet industry needs [3]. In recent years, online learning platforms and massive open online courses (MOOCs) have become pervasive for workforce training due to the flexibility in time and location as well as the capability of learning analytics to support the analysis, visualization, communication, and management of learning processes [6, 7]. Additionally, online courses and brief topical certificates reduce costs regarding physical materials, spaces, and allow for economics of scale, one certificate can teach critical competencies to thousands of learners. MOOCs promote (a)synchronous learning in which learners are able to follow their own paths to learn the content with sequences of learning modules that enable learners to advance personalized learning [8].

Educational data mining (EDM) and learning analytics (LA) as a set of emerging practices have been used to examine learners' trajectories, activities, proficiencies and provide useful information for instructors to understand employees' learning [9]. While these EDM and LA practices and approaches are used commonly in educational settings, few studies examine how they are applied in workplace settings [8]. To bridge this gap, various fields have seen an emerging trend where companies and educational institutions work collaboratively to develop courses to meet specific stakeholder goals. While these efforts are laudable, they typically do not benefit from the latest understandings from the learning sciences and do not use data visualization to scale high-quality learning outcomes to the gamut of online skills development offerings required by rapidly accelerating S&T developments.

This study illustrates the potential of applying learning design to a course focused on upskilling of employees of a large aerospace manufacturer, offered via MIT xPRO's edX platform. The MIT xPRO Additive Manufacturing (AM) Certificate represents a new credential that blends industry expertise with traditional academic learning. The collaboration between industry and academic partners in this effort was designed to find a nexus between the learning sciences and engineering education research, uncovering how to improve workplace learning outcomes through the scientific process of exploration, discovery, confirmation, and dissemination. The certificate case studies were created with industry experts and industry TAs were enlisted to provide feedback throughout the learning process. The goal of this research is to contextualize evaluation metrics by taking into consideration learner actions, course characteristics, and stakeholder goals. Study results include data collection instruments and an evaluation method that makes sense of diverse forms of learner data for action-oriented decisions to support learning and institutional goals, for improvement of the AM course and guidance of online learning initiatives more broadly.

## Background

**MOOCs and Workforce Training**

Even outside of pandemic times with stay-at-home orders, many companies value, purchase, and promote online education and upskilling. Professional development and workplace training have offered targeted educational experiences and just-in-time resources for years [3]. While traditional Instructor Led Training (ILT) is considered costly and has certain limitations in terms of scaling company expertise, web-based training is seen as a more cost-efficient alternative as it offers selectable study times, self-paced instruction, and doesn't require the use of physical spaces [10].

As an example of online training, Massive Open Online Courses (MOOCs) are a form of online learning designed for unlimited participation and open access via the internet. MOOCs are traditionally developed by universities and other educational providers in an effort to support lifelong network learning as well as create open, participatory, and distributed learning spaces [11]. In addition to the hosting of asynchronous course materials, such as filmed lectures and readings, MOOCs often provide interactive elements in the form of user forums and online tools that offer instant assessment feedback to students.

The emergence of MOOCs has evoked a new wave of research and expands the frontier of digital and online learning [12]. Particularly, the implementation of micro-credentials as a certification of skills attained through MOOCs has captured the interest of universities and workplaces as they provide ways to measure the experiences of continuing education students, nontraditional learners, or employees rising through ranks within an organization [13].

One of the leading MOOC providers is edX, a nonprofit education provider hosting online courses from universities around the world across a range of disciplines. Courses run on the edX open-source software platform often involve multiple module types (e.g., html, video, or discussion forum). Courses hosted on the edX platform are organized into chapters, each involving a sequence of videos, content pages, and customizable modules, including problem questions, open assessments, interactive visualizations, or discussions; this structure was retained for the MIT xPRO additive manufacturing course. Specifically, MIT xPRO courses are offered with fixed enrollment periods. More broadly, many edX learning experiences, of which MIT xPRO's offerings are a subset, combine to create their own form of micro-credentials (e.g., MicroMasters® programs).

Like many online course providers, the MIT xPRO organization is interested in supporting research on online pedagogy and learning. However, most existing research on MOOC pedagogy and learning are limited to academic fields, particularly in higher education [14]. Given the lack of empirical evidence to understand how MOOCs are used in industry and corporate training, this study aims to expand the body of research on the implementation of MOOCs in engineering workforce training by examining how engineers learn the topic of additive manufacturing in an online course offered through MIT xPRO's implementation of the edX platform.

**New Digital Fabrication Technologies**

The introduction and adoption of interconnected digital systems are changing the nature of work across industries and within organizational roles. Consequently, traditional educational paradigms must evolve and scale the instruction of emergent workforce competencies, both in terms of their time requirements and content alignment with key learning objectives. Moreover, advances in fields such as artificial intelligence (AI) and robotics are making it increasingly possible for machines to perform not only physical but also cognitive tasks currently performed by humans. This will result in relationships where humans and machines work collaboratively, rather than in a command-and-control relationship, suggesting both the opportunity for new working modes and the impetus for novel instructional methods to meet these new requirements. These trends are reflective of broader changes occurring in society around the exchange and acceleration of knowledge, including information enabled through or by AI and machine learning (ML).

In manufacturing and engineering industries, several key technologies that will impact the workforces' knowledge development and learning needs. These include: robotics and automation; industrial internet of things and advanced sensing; digital twins (corresponding data packages for each design and part); cyber security for production systems; augmented and virtual reality, e.g. for inspection tasks; and additive manufacturing (AM) technologies; among others (e.g., [15-17]). Taken in total, these technologies are posed to transform the design and capabilities of complex manufacturing systems, enabling greater operational flexibility (in the forms of reconfigurable production assets) and strategic flexibility (in the form of granular, data-driven business intelligence). In exchange, companies must commit significant resources towards system integration and human capital development, and therefore it will be ever more important to develop the skills of working individuals to engage in creative knowledge work and make sense of increasing complexity as these technologies mature and intersect.

One important example of technology-driven organizational change is the case of additive manufacturing (AM), which is alternatively known as 3D printing. The AM industry has grown rapidly in recent years, and the global AM market for machines and services alone has grown rapidly. In 2019, the industry was sized at US$9.79 billion and is buoyed by aggressive growth; the 30-year compound annual growth rate (CAGR) is measured at 26.9%, with a growth rate of 33.5% in 2018 well-above the historical mean [18]. Importantly, the use of AM has expanded from rapid prototyping, where it has been used since at least the early 1980's, to

manufacturing applications, including for volume production of end-use aerospace and medical components. AM is considered a revolutionary technology that has will change the design and production approaches of manufacturing firms [19]. Compared to traditional manufacturing processes such as casting or machining, AM is unique in that it does not required part-specific production tooling and thereby enables greater design flexibility, resulting in possible advantages in shape-optimization, time-to-market, waste reduction, assembly and supply-chain consolidation, and more [20].

Given AM's recent introduction to volume production contexts, most engineers and manufacturing workers were not exposed to, or trained with, the principles or execution of AM during their formal education. It is likewise challenging and time consuming for universities to construct new degree programs and commensurate curriculum in AM [21]. Additionally, the use of AM as a forming process does not necessarily change the typical considerations when designing a given component. Rather, AM is a new approach to forming a part to fulfill a pre-established function, and the convergence of "gray-hair" industry- or application-specific knowledge with AM-specific knowledge is necessary for the technology to be best utilized in manufacturing environments. Finally, the nascency of AM in many production use-cases, coupled with the rapid growth of industrial interest, has resulted in an industry prone to rapid changes in technological capabilities and demonstrated use-cases. Therefore, professional training in AM is a key enabler to accelerate the learning process to meet the current needs and demand in the workforce. High-quality courses and instruction in AM are needed to fulfill this requirement for training skilled employees. In this study, the aerospace company worked with faculty, learning scientists, and instructional designers to develop an online Additive

Manufacturing course that can be iterated upon to keep pace with rapidly changing industry needs, both in content and in structure.

**Learning Objectives and Cognitive Load**

One of the important starting points when iteratively designing a course with the learning sciences in mind (Barab & Squire, 2004) is to begin by articulating desired learning objectives [22] and backward-designing [23] course activities and assessments to more closely map student activity toward achieving and assessing those objectives. Learning objectives are goals that target specific knowledge, skills, and dispositions taught in specific sections of a course [22]. The revised Bloom's taxonomy (see Table 1), a hierarchical model used to classify educational learning objectives into levels of complexity and specificity, is commonly used by instructors to design and develop their own learning objectives across a range of subjects [24]. The taxonomy differentiates between cognitive skill levels, emphasizing the hierarchical relationships from lower levels (less complex, less specific) to higher levels (more complex, more specific). These cognitive skill levels are ordered into six categories: 1. Remember, 2. Understand, 3. Apply, 4. Analyze, 5. Evaluate, and 6. Create. These levels are adapted by educators as an instructional tool to design and develop learning objectives [24]. By identifying these processes as operative nouns and verbs of a given learning outcome, instructors are able to develop an assessment to evaluate specific learning constructs and skills. In ideal learning environments, learning objectives include higher levels of cognitive skills for learners to lead to deeper learning and knowledge transfer to a variety of contexts [22].

Relatedly, Cognitive Load Theory provides educators and instructors a way to develop instructional procedures by keeping in mind "aspects of human cognitive architecture that are relevant to instruction along with the instructional consequences that flow from the architecture" [25] (p.6). This theory explains that learning is a serial process of information from working memory to long-term memory. Central to Cognitive Load Theory is the idea that ideal learning occurs when instructors effectively guide learners from lower cognitive level activities (e.g., remember factual knowledge) to higher cognitive level activities (e.g., create new knowledge).

**Table 1. The Revised Bloom's Taxonomy [24].**

| Cognitive Skill Level | Description |
| --- | --- |
| 1. Remember | Retrieve relevant knowledge from long-term memory. |
| 2. Understand | Construct meaning from instructional messages, including oral, written, and graphic communication. |
| 3. Apply | Carry out or use a procedure in a given situation. |
| 4. Analyze | Break material into constituent parts and determine how parts relate to one another and to an overall structure or purpose. |
| 5. Evaluate | Make judgments based on criteria and standards. |
| 6. Create | Put elements together to form a coherent or functional whole; re-organize elements into a new pattern or structure. |

To understand the cognitive load levels of the MIT xPRO AM course, this study applied the revised Bloom's taxonomy to analyze the course content and assessments. Each learning activity and assessment was associated ("tagged") with categories from Bloom's taxonomy and cross-referenced with analyses of the course's desired learning objectives. Ideally, each learning objective should be supported by a gradual progression through the process category hierarchy of the Revised Bloom's Taxonomy, where each objective was taught by first introducing the learner to low-cognitive-load tasks and increasing over the length of the course in terms of complexity and cognitive sophistication. This paper advances this work to offer strategies to visualize and compare this ideal with how learning objectives were implemented in the course.

**Visual Learning Analytics**

To ascertain the mechanisms of student performance in the course, a range of learning analytics (LA) methods were employed. LA approaches are widely applied to understanding learning in online and web-based learning [26, 27]. Learning analytics is an interdisciplinary field which includes the fields of artificial intelligence (AI), statistical analysis, data mining, learning sciences, and machine learning [28, 29]. In learning sciences, learning analytics offers various approaches to analyze and interpret learners' data, such as statistics and visualization, data mining, prediction, clustering, and relationship mining regarding the applications and purposes of learning [30]. The implementation of LA provides effective information from a volume of learners in a given population over time to identify learning patterns or learner trajectories.

Instructors recognize the potential in using LA to personalize learning, inform practical experiences, and provide self-regulated learning opportunities. Particularly, the application of LA creates different views of learner activity data which instructors can use to track learners' performance and styles, training outcomes and histories, and community of practice participation [21]. Building on the approaches of LA, visualization methods help instructors to understand and optimize student learning as well as inform their iterative decision-making processes in online learning [31].

While LA is an effective tool for instructors to make sense of diverse student data, it can also directly benefit the learners who are given the tools to quantify and analyze their own performance. Building on the implementation of LA in MOOCs, visual learning analytics is a discipline aiming to assistant learners to interpret complex user data. Visual LA is defined as "the science of analytical reasoning facilitated by interactive visual interfaces" (p.4) [32], which integrates areas of user interaction, data science, and visual representation. It provides interactive visualizations to help users in sense and decision-making processes in learning. Because it bridges the connections between users and their data, instructors, administrators, and users themselves can deal with data-related tasks more efficiently and effectively [33]. Although visual analytics tools have been employed successfully in studying users' data in MOOC platforms (e.g., exploring learning paths or progress) [34, 35], improving instructional design [36, 37] and understanding peer collaboration [38], there is a lack of studies that "employ sophisticated visualizations and engage deeply with educational theories" [33] (p.1).

This study applies a visual learning analytics approach [8, 39, 40] to extend the focus of learning analytics and apply visualization techniques to examine teaching and learning in workforce training via analysis of the aforementioned AM course. Using systematized data structures and visual learning analytic dashboard designs, we link learning objectives analysis, learner trajectories, engagement, and performance in order to examine the efficiency of learning and to provide suggestions to course designers and instructors in online courses. Beginning with a central research question which aims to assess how visual learning analytics may be deployed to support instructors in online course design, we examined the following questions:

a. To what extent is engagement with course content related to overall student performance in the course?

b. To what extent do the course learning assessment measures map onto the learning objectives?

c. What is the relationship between time spent, student performance, and the identified learning objectives?

d. To what extent does the Bloom's Taxonomy analysis reveal areas to iteratively improve in subsequent course design?

# Method

A nine-week online Additive Manufacturing course was developed in 2018 for the MIT xPRO edX platform, via collaboration between experts in AM and employers in industry. See Appendix A for a high-level overview of the course that outlines the topics of lessons, expected completion time, and course schedule. This study focuses on weeks one to six because in weeks seven to nine, employees used a 3D CAD modeling design platform that generates proprietary data exclusively and little information exists on the course site regarding learning activities and performance for these three weeks.

## Participants

The course enrolled a total of 930 individuals in engineering and manufacturing-related roles, all employed by a single, major aerospace company. Participants had a range of work experience (ranging from little professional experience to more than 15 years) and educational backgrounds (ranging from a high-school degree or equivalent to PhD). The demographic information was gathered from an entrance survey required for all participants who took the course. Pre-assessment was required for all participants before taking the course.

## Data and Analysis

Before the 2018 deployment of the course, the company engaged in a two-prong approach to evaluating the course: by enlisting a group of employee beta-testers that reviewed a beta-release of the course as a professional development opportunity and provide learner feedback to the course team; and with a group of learning scientists and data visualization experts to look closely at employees' learning outcomes, engagement, and trajectories within the course. The objective was to analyze a range of data related to course deployment and usage to identify areas of success (or areas to improve) within the training.

**MIT xPRO data**

Data from an MIT xPRO course comprises a (1) course database, which captures **course structure**, (2) **learner demographics and performance data**; and (3) daily **event logs**. All three data types are briefly described here; details can be found in the edX documentation [28].

**MIT xPRO course database.** *Course Structure*. The hierarchical structure of learning modules exists in JSON format. Modules can be of different types such as html, problem, video, or discussion. The five-level JSON tree hierarchy and associated sequence of learning modules, together with a one-dimensional base map of course modules, is organized linearly from beginning (left) to end (right) according to instructor design. The root node provides course metadata; the second level corresponds to course chapters that outline the major groupings of content; the third and fourth levels represent learning modules; and the fifth level nodes are course content blocks (e.g., videos, html pages, problems, assessments, etc.). Module identifiers from various levels of the course structure are linked to the course clickstream event logs (see Data Types and Descriptions) through a variety of references specific to the type of module and action recorded.

*Learner Demographic and Performance Data*. Learner data, captured in a SQL database, includes learner-generated artifacts, system and social interactions, and learning outcomes. Among others, the data captures first and last learning module states, final course grade, certificate status, discussion forum posts, and team data. Discussion activity for the course was supported via a native discussion forum endemic to MIT xPRO's edX installation but cannot be linked to other activity data and hence is not used in this study.

**MIT xPRO event logs.** *Event Logs*. Learners' interactions (along with course developers' and instructors' interactions) in a course are emitted as browser and mobile events, and the learning management system's responses to user interactions are emitted as server events. Both are available in real-time as streaming JSON (ndjson) records. Logs of learner activity are captured daily for each MIT xPRO course. The logs for the AM course contain events associated with content knowledge (videos and html-text modules), graded problems (e.g., multiple choice or select-all-that-apply problems), and open response assessments (see supplemental materials for details).

## Learning Objectives: Coding and Analysis

Learning modules were qualitatively coded with a total of 31 learning objectives (LOs) to examine if these LOs and associated AM skills were well aligned with the course content. Learning objectives include both the acquisition of knowledge and the application of that knowledge in the form of discrete skills. For the purposes of this paper, each LO was assumed to be equally important relative to one-another and therefore, in an ideal course deployment which matches this assumption, would receive equal and sufficient attention in instruction and assessment. In the future, one could assign differential weighting to LOs deemed more central to the objectives of the course versus those which are peripheral or less significant. The LOs were originally articulated by the course instructors and were later refined by the learning sciences research team. Analyses included tagging the 983 course activities (including videos, readings, and course assignments) with one or more of the 31 learning objectives and skills of the course. The goal of this qualitative coding was to visualize and analyze the relationships among the content of course modules, assignments, and learner performance.

Several learning objectives were covered during each week of the course and throughout each learning module. Each of learning modules was coded for up to three of the most pertinent LOs within the activity. This allowed for the course designers to more granularly look at course outcomes to better understand whether students were (un)able to demonstrate a particular learning objective due to inadequate course materials or insufficient time within the course spent on that learning objective (see S1 Supporting Information).

## Cognitive Load: Coding and Analysis

For the purposes of examining how increasing complexity within the course related to learners' cognitive load and outcomes, a second set of codes was applied to the learning modules following the revised Bloom's Taxonomy [24]. Table 1 illustrates the six categories. Each course activity (e.g., a video, assignment or page) was qualitatively identified as one of the six categories according to the revised Bloom's Taxonomy. Additionally, the frequency and percentage count of each category within the taxonomy were calculated to demonstrate the distribution of cognitive load by course week.

The application of Bloom's-informed codes enabled instructors to evaluate whether they considered cognitive load in the sequence of course activities and time spent. In short, this framing would advise that (a) we would see a full range of activities integrated into the course design, and (b) we would expect early units in the course to engage in less-challenging process categories and would build toward more higher-level tasks from a cognitive load perspective. Ultimately, the results of the qualitative coding were visualized to cross-compare student's dwell times and grade performance against the course's learning objectives (Figs 1 & 2). Data processing and visual analysis was completed using R statistical software and packages. The purpose of the analyses was to examine if the content of the course modules and assignments were well designed to evaluate employees' learning which can be reflected in their performance.

Student learning objective dwell time statistics are calculated using student event logs that were processed using the MIT xPRO learning analytics pipeline [8]. Processed student event logs include content module identifiers and dwell time calculations based on the temporal differences between student activities, where periods of time over 10 minutes are considered breaks in strings of activities. Individual student event logs are joined to content modules and learning objective codes and then aggregated to count the number of events and sum dwell time associated with each learning objective.

To evaluate student performance for the course's stated LOs, this study relies on course subsection grades calculated by the MIT xPRO system for each student. Subsection grades are captured under sequential modules from a given course chapter, and include a student's identifier, points earned by a student and the points possible for a sequence of content modules. Sequential modules without points associated with content were removed from the data set. These data were then aggregated by learning objective and student identifiers to calculate the total points associated with a learning objective, the total points and percentage earned by a student for a learning objective.

# Results

To understand how the instructional design supported employees' learning, the study examined the learning modules with cognitive load analysis, and then applied visual analytics and qualitative coding to investigate the relationships between learning objectives, performance, and engagement.

## Engagement in Course Content

Overall, our analyses confirm that time spent (i.e., engagement) was positively correlated to student performance (r =.56, p < .01, n = 930). While this is not surprising, it is an important measure to run at the end of any course to ensure that students, regardless of background knowledge, can perform well if they invest an adequate amount of time in the course.

Examining the intersection of student performance, assessment measures, and intended learning outcomes yielded additional insights into how students engaged with the content. Taking a closer look at the amount of time spent on each learning objective (which might not necessarily have a 1:1 correspondence with a particular activity or assessment) has two primary benefits. First, by providing visual learning data analytics dashboards, where instructors can look at average time spent by learning objective, this analysis helps to establish where students are spending their time in the course content, as well as establish more accurate estimations of time toward completion of each module. Second, the visual LA can clarify the amount of time it should take for students to accomplish a specified learning objective or task.

Fig 1 presents the time students spent on each learning objective. This view makes salient how much time proportionately is dedicated to a learning objective, either in terms of time each student spent to complete a task, or how much extra time is dedicated in terms of learning activities to particular learning objectives. When we compare time spent per learning objective against student assessment outcomes by learning objective (Fig 2), we see that generally everyone scored high in the first set of learning objectives, which also required a small amount of time to complete.

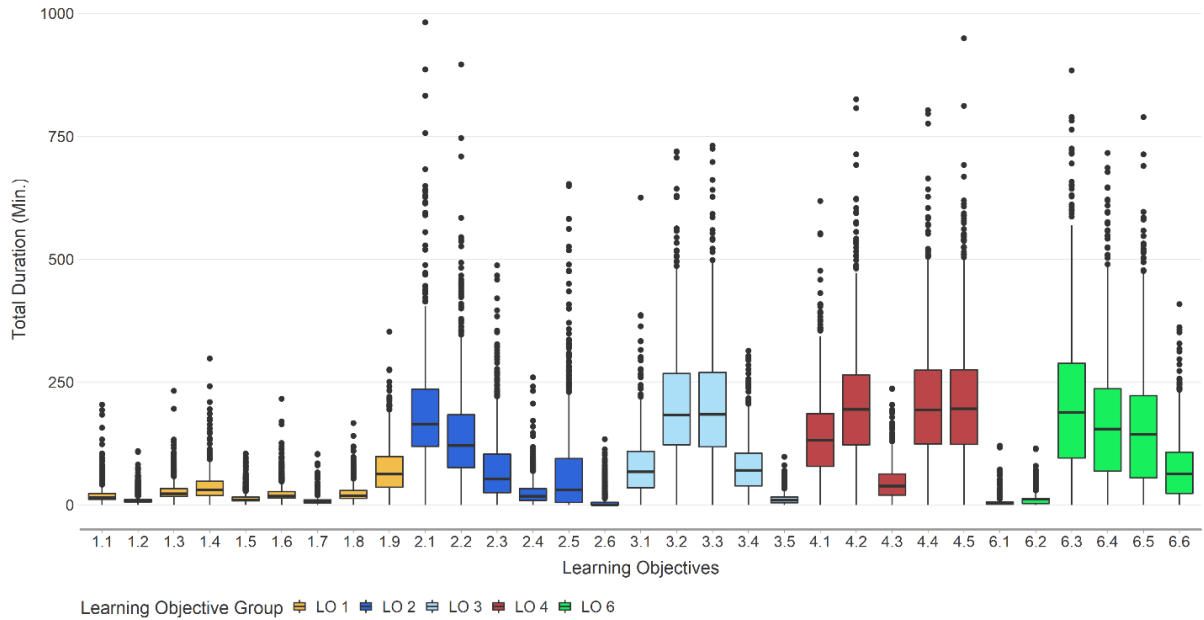

**Fig. 1 Engagement Time by Learning Objective.**

Fig 2 illustrates the relationships between the learning outcomes and learning objectives. The learning objectives are numbered, and color coded by learning objective groups, see Fig 3 for an alignment of six weeks and learning objectives.

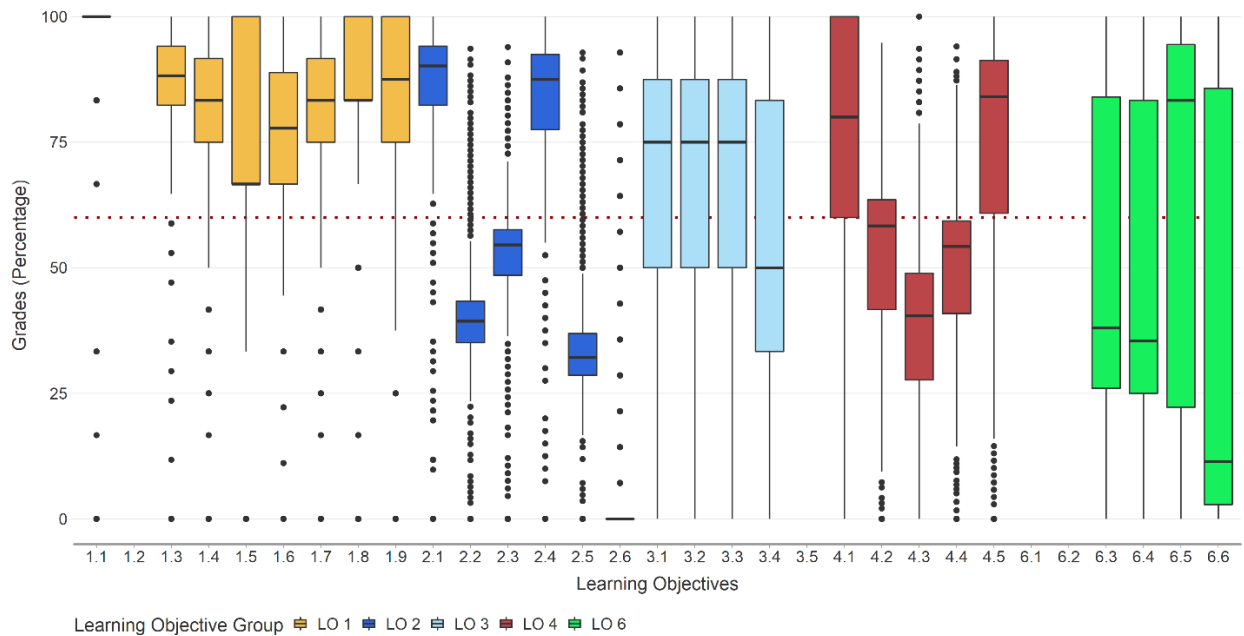

**Fig 2. Box plot of percentage grades for all students on learning objectives covered.**

Comparing these results in the visualization dashboard suggests several things to a course instructor: first, the LO1 learning objectives group may be relatively easy to achieve in a short period of time, and so less time could be allotted to that course content. Second, students coming into the course may be already familiar with these objectives, so this unit could be modified to better account for the background of the learners. For example, effective pre-assessment could be used a selection tool to differentiate student populations which require this material as a pre-requisite to more advanced material from those populations which may bypass or "test-out" of the requirement. Alternatively, content could be modified based on learner demographics and background to better suit gaps within existing knowledge. These dashboards are intended to have instructors ask questions of their course based on time spent and outcomes achieved to make changes that best fit the needs of employees and businesses.

In addition, oftentimes, instructors (as was the case in this particular course) are asked to state the amount of time it would take to complete each unit, often relying on rough estimates the instructor thinks it should take to complete a task. Frequently, this can lead to student frustration when a particular unit is disproportionately heavy in terms of time commitment. Instead of instructors estimating these numbers, the visualizations show the actual amount of time spent--retrieved from the data logs. Instructors can use this information to consider how activities can be shifted between modules to create more balanced workloads, or in the design of narrowly tailored learning objectives. In Fig 1, LO2.1 is associated with significantly more time-spent than LO2.2 and those thereafter; suggesting that, for the purposes of analysis and iteration, the objective, LO2.1, may be more effectively divided into a series of discrete objectives for purposes of instruction and assessing learning.

Furthermore, analyzing overall distribution of student performance within each LO and longitudinally can yield additional insights. For instance, for instances where certain LOs produce very low scores, instructors can look for correlation between performance and time spent. If more time spent on a LO leads to a low score, instructors can infer that students are putting forth effort, but the course content could be improved in future iterations.

## Learning Objectives and Assessment Mapping

By mapping LOs to student performance, instructors are able to determine areas where students may have consistently struggled, thereby identifying areas to inform future iterations. The visualization in Fig 2, which aligns LOs by assessment measure, reveals that some desired learning objectives lacked a corresponding assessment measure to test whether that objective was achieved through course participation. In other cases, there may be too many measures to quantify a particular objective, to the assessment detriment of other objectives of the course.

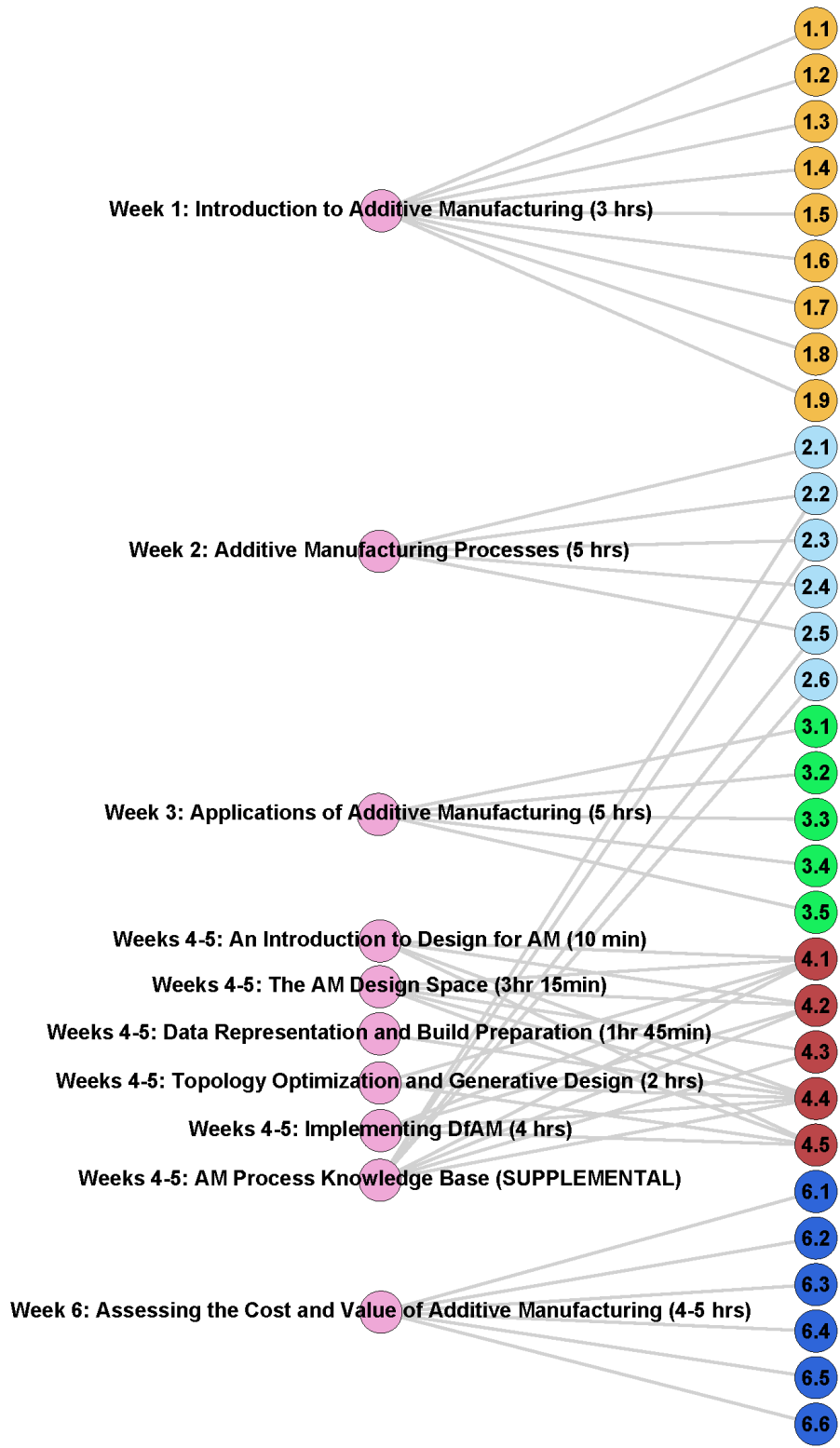

**Fig 3.** Relationship between course modules (left) and learning objectives (right).

In Fig 3, each discrete content element within a given module or week (e.g., chapters, pages, and the content within them - videos, html, problems, and so forth) were aggregated into a top-level categorization shown by the monochromatic circles on the left. On the right, each circle corresponds to an LO. The mapping between these two illustrates the relationship between course construction and attention given to each LO. In week 1, the number of modules was average and equally covered each LO in the week (see Fig 3). In week 2 and week 3, the emphases of LOs were not evenly distributed across the corresponding modules. Particularly, LO 2.2 and 2.5 were highly emphasized in week 2; LO 3.2 and 3.3 were highly emphasized in week 3. Week 4 and 5 had the highest number of modules across the course, and some of the content was connected back to the LOs for week 2. The content of modules in week 6 was addressed evenly across LOs in the week. This is not entirely unexpected; the course was intentionally designed to scaffold the presentation of concepts and principles which, at first glance, appear disconnected and unrelated (e.g. teaching the use of an advanced design software tool versus using cost-calculation methods to estimate component cost). As the course progresses, these disconnected principles are gradually re-introduced through their relationships on other concepts and principles in the course. Arguably, this design choice illustrates a limitation of the current study methodology, since the study assumes that LOs are canonically presented linearly and ought to be assigned equal attention, and thus the AM course's construction was misaligned. Conversely, this nuance elevates an important point in the use of visual LAs for course optimization, insofar as the evaluation of a courses' approach must be designed in lock-step with course design to generate actionable, context-dependent recommendations.

Another way to view this visualization is to interrogate areas where students collectively score low. Cases such as these might suggest that there is a misalignment between assessment measure and learning outcome, especially when students are shown to spend a good amount of time on content teaching that objective. An example, seen in Figs 1 and 2, is where students struggled with LO group 2, though students spent a disproportionately large amount of time on activities teaching those particular objectives. Again, however, this may equally reflect the significance of designing narrowly-tailored learning objectives; in Fig 3, the LO2 group is associated both with Week 2 material and material located within a "Supplementary Knowledge Base." The "Supplemental" material is a series of optional, ungraded content comprising many hours of rigorous technical instruction; the amount of content is a multiple of that presented in Week 2, and it is also the most technically advanced and challenging material in the course. The "Supplemental" section is also associated with the LO4 group, which received the second-most attention of any LO grouping behind the LO2 group (Fig 2), lending credibility to this hypothesis.

Moreover, the LO2 category in this example is practically multiplicative; each LO within the LO2 cluster is associated seven times with an individual AM process. An LO written to say, for example, "Understand the mechanical properties of parts produced by additive manufacturing," is actually applied to each of seven AM processes addressed within the LO2 cluster, and thus could be authored more precisely seven unique times for each LO in the following syntax: "Understand the mechanical properties of parts produced *for process 1 (2, 3, ...)*." In this case, the convenience of writing broad, truncated LOs trades-off directly with the utility of visual LAs to provide course insights, and therefore reinforces the necessity to align retrospective evaluation methods up-front with course design.

As shown in the Fig 2, not all objectives were assessed, while some objectives possessing a disproportionate amount of assessment. The learning objectives on which students consistently scored high or low indicate the lack of corresponding assessment measures which efficiently evaluate the course content. For instance, the results of learning objective 2 (LO2) (blue bar in Fig 2) showed that the course content might not well align with the assessments in week 2, as both passing and not-passing employee groups showed low achievement on most of the learning objectives in that week. To extend the analysis further, for the two lowest-scoring LOs within LO2, LO2.2 and LO2.5, significant instruction towards these LOs is located within the "Supplemental" material for the course. This helps explain both why these LOs commanded significant time, insofar as they were associated with greater amounts of instructional content, and why student outcomes were comparatively low, insofar as these LOs were associated with limited grading during required modules and no graded assessments during supplemental modules.

Taking these LOs as an example, the results allow instructors to know where and when students got hung up on the course content or assessment, and provide initial insights for course designers to revise and modify the learning modules and objectives. In this case, a next revision of the course may include demarcating LO2 into several learning objective categories, where those taught and assessed during required content are differentiated from those primarily taught and assessed via supplemental content. This differentiation may also occur within each LO to better map objectives against lecture material, rather than aggregating individualized LOs into broader categories (as was done with LO2), to clarify the practical implications of analysis and design more effective, targeted assessments.

In short, this result may indicate that the assessment measure may be misaligned with how the concept was taught, or it could also mean that those course materials were insufficient to address those goals. It is important to underscore that these visualizations are designed to empower the instructor to make sense of this data and to look more closely at their students and materials to ascertain why these outcomes were reached. This should be a normal part of course development cycles. Those in the learning sciences would routinely think about and toggle between these understandings in high-quality curriculum design, while instructors may not generally be equipped to think about course design and learning objectives in this manner. Therefore, it is hoped that these visualizations help examine and improve the alignment between clear learning objectives, high-quality course materials, and high-quality assessment measures so all learning objectives are met.

## Cognitive Load Results

For each week of the course, materials were tagged with process categories drawn from the revised Bloom's taxonomy [24], depicted above in Table 2 and visualized in Fig 4. Fig 4 reveals how time spent and the types of processes engaged were analyzed for each week of the course. Cross-referencing Fig 4 against the learning objectives depicted in Figs 1-3 indicates which learning objectives were taught using a robust range of process categories, versus others that were assessed using primarily lower-level processing.

As shown in Fig 4, the process of comprehension/understanding was heavily emphasized from week 1 to week 6. Conversely, higher cognitive load processes (i.e., applying, analyzing, evaluating, and creating) were not introduced until weeks 7-8. Most of the assessment in weeks 1-6 focused on the first and second levels of the revised

 Bloom's taxonomy. Visualized in this way, instructors can revisit their assessment means to introduce a fuller range of activity earlier in the course, or to produce a more gradual complexity of the assessment means over time. This analysis helps us to further determine which weeks of the course were given a range of high-level processing, and whether increasing complexity was asked of the learner over the span of the course. Importantly, Weeks 7-8 are comprised of a two-week long "case study" assignment; they do not include instructional materials but rather are spent completing an assignment. Unfortunately, due to aforementioned reasons relating to data fidelity, these weeks are not addressed in Figs 1-3.

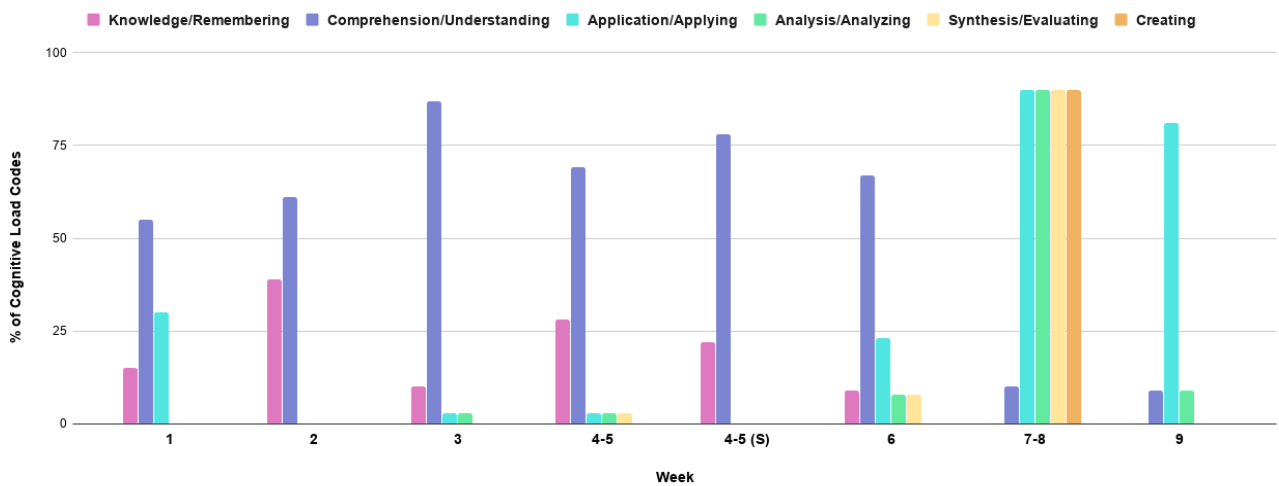

**Fig 4. Cognitive Load Results of Additive Manufacturing Course.**

From this figure, it is evident that while the course did scaffold tasks of increasing complexity over the course's duration, this scaffolding occurred linearly *across* LOs (each LO cluster corresponds to an associated week, identified by the same numbering scheme; i.e., LO1 corresponds to Week 1 of the program) rather than *within* a given LO cluster, with some exception to LO6. A future course may instead structure tasks and assessments of increasing complexity within a given module to ensure that students "climb" the taxonomy for each LO grouping, rather than only in subsequent LOs. Practically, instructors may implement this proposed approach by evaluating their assessments within each LO, as it may be possible to transition assignments that are aligned with lower levels of the revised Bloom's taxonomy to higher levels through careful design of required student activities with modification.

Importantly, the analysis elucidated via Fig 4 weights each data point equally; in other words, a multiple-choice question aimed at assessing understanding (which may take several minutes to complete) is treated equally as an open-response question aimed at requiring synthesis or analysis (which may take an hour or longer to complete). It is therefore possible that certain cognitive loads are under- or over-represented in the figure. Future work may instead assign time-associated weighting criteria to each data point to clarify identified discrepancies between idealized course construction and actual course construction for the purposes of improvement. However, in the minimum case, this data is still useful as one would expect to see a distribution of activities (irrespective of their frequency) across Bloom's taxonomy, which was not evident for each LO and week evaluated.

## Discussion

This study examined the learning processes and outcomes of participants through an online course which focused on the topic of additive manufacturing. Based on the analyses of learning objectives, processes and outcomes, the results informed potential modifications of future course content and assessments. This process could be applied to courses on many other topics. MOOC platforms could integrate simple supporting workflows aimed to articulate learning objectives and tag them in the course assignments and assessment metrics. These components and functions of the platform could address the course design challenges listed above and improve the teaching and learning processes for future courses. Although the data and analyses were drawn from a single course, these provided a springboard to further the landscape in using LA across online courses and platforms and data structure for the purposes of adaptive assessment.

Further, visualizing employees' performance by learning objectives illuminates where and when learners had challenges on course content and assessment. Examining the intersection of student performance, assessment measures, and intended learning outcomes yielded additional insight into the effectiveness of the training. In sum, five key points are emphasized to guide design of online courses.

1. **Start by defining course goals, objectives, and outcomes**: Start by defining your expectations for what students learn in the course, rather than by detailing the content that your course will cover.

    a. Course goals: High-level items students should acquire during your course. Be as specific as you can and make sure that the goals define learning in ways that can be measured.

    b. Learning objectives: Write brief statements for each section of the course that describe what students will be expected to learn in order to proceed. Good learning objectives are meant to be broader learning goals targeting important knowledge, skills and dispositions taught in that section of the course. Be willing to revise learning objectives based on results if it seems there are too many measures to quantify a particular objective, to the assessment detriment of other objectives of the course. Researchers may also want to consider the hierarchical relationship between the LOs (if any), as well as relative weighting of the objectives that would have implication for time spent and frequency. As this study suggests, careful design of LOs in concert with design of LA evaluation during initial curriculum development may yield more practically useful information than if the two activities are performed asynchronously.

c. Outcomes: Statements of knowledge or skills students should acquire within a particular assignment. This is where you define the content your course will cover. Outcomes can be a comprehensive listing of discrete skills or bits of knowledge you teach at each point in the course.

2. **Construct high-quality learning through research-driven instructional systems design:** Upskilling is not the accumulation of individual knowledge but rather positioning the learner to create and apply collective and applied procedural knowledge. Courses should develop pathways where learners are tasked, wherever possible, to apply knowledge with a mix of expertise (e.g., aerospace engineers apply AM skills through design and manufacturing activities) thus deepening and evolving conceptual understandings within a relevant practical context [41, 42]. This is the basic path from novice to expert, and is especially relevant for the field of AM.

3. **Consider your student population:** Understand the students who typically take the course in order to think about how your course will help this group of students build their knowledge and understanding of the topic.

    a. Level of preparation or interest: Full-time college students have more time and resources available to them than people who are working full time and taking a course for additional knowledge or certification, and may be working in their spare time such as during evenings or weekends.

    b. Amount of time for working on course content: Compare the time which students are expected to spend in the course and the actual time they spend. This can better help students to reflect their own learning trajectories and progress in specific content areas.

c.  Level of skill and interest: Use surveys or exercises to identify students' prior knowledge and interest areas. Use and build upon existing knowledge and interests which allows course materials to better focus to engage learners.

4. **Determine how you will evaluate learning:** Assessment must go hand-in-hand with course goals and represent the revised Bloom's taxonomy that you've outlined for the content. For example, if the goal is to improve problem-solving skills, the exam (as well as any assignments leading up to it) should contain questions that ask students to recall facts as well as specific problem-solving exercises. If one form of assessment dominates the activities throughout the course, be open to revising assignments in order to elevate them to higher order thinking as the course progresses (e.g., incorporating more opportunities for strategy formation, solution monitoring, and creativity to activate higher stages of the Bloom's Taxonomy).

5. **Adapt and rework the course dynamically:** Instructors should monitor student progress and make adaptations as needed. If you see that a large percentage of your students are not accessing resources, participating in your community of practice, or completing assignments in a timely manner, be prepared to make changes. Be willing to revise, and in some cases, abandon course practices and content based on feedback from learners and from learning analytics. In the case of the AM program, course instructors created a detailed exit survey which asked students to reflect on the duration, technical rigor, and constitution of each LO grouping as well as the overall course presentation. Fig 5 shows one such question, which asked students to evaluate a series of possible modifications to the program; students indicated strongly that the overall duration of the program should be increased, and that, due to the breadth of content addressed (a principle reinforced by the asymmetry of dwell-time and student outcomes for certain LOs), more granular or modular course offerings should be presented in the future. These principles were adopted by instructors; the duration of the course was increased to 12 weeks (the most popular response to a separate question dedicated to overall course length), and additional modularization is planned.

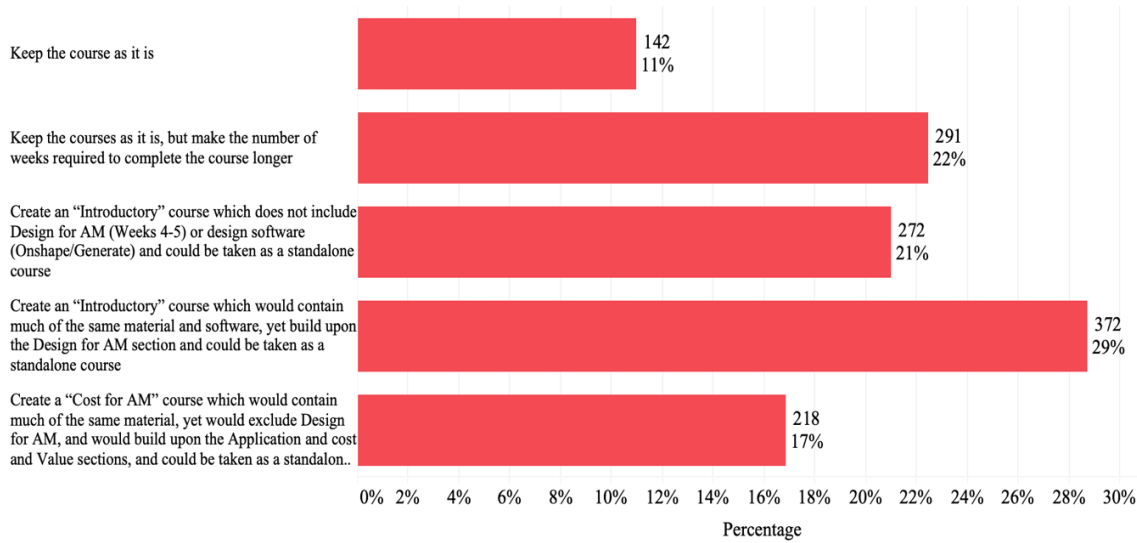

**Fig 5. Exit Survey results to the question, "Which of the following options would you support as strong alternatives for changing the model of how the course was delivered? Select all that apply."**

In summary, the process of upskilling employees in emerging technologies such as AM requires close collaboration of learning scientists, data scientists, security experts, learning platform developers, and industry to design and develop learning materials and environments which lead to measurable improvements in work performance. This study shows how learning sciences can improve the design and effectiveness of a large-scale online course delivered to a broad technical audience, and how performance can be assessed from analytics of learner engagement and performance. Particularly, the data structures and visualizations inform strategies for course instructors to improve alignment of course content, assessment measures, and learning objectives.

# Acknowledgments

[Do not include funding or competing interests information in Acknowledgments.]

**References    Uncategorized References**